\documentclass[11pt]{article}

\newcommand{\beq}{\begin{equation}}
\newcommand{\eeq}{\end{equation}}
\newcommand{\psib}{\ensuremath{\overline{\psi}}}
\usepackage{latexsym}
\usepackage{graphicx}

\begin{document}
\begin{flushright}
SU-4240-721\\
\today
\end{flushright}
\vspace{0.5in}

% declarations for front matter
\begin{center}
{\Large\bf A lattice path integral for supersymmetric quantum mechanics}

\vspace{0.5in}
{\small Simon Catterall$^*$ and Eric Gregory$^{\dagger}$\\
        $^*$ Physics Department, 
        Syracuse University, 
        Syracuse, NY 13244 \\
        $^{\dagger}$
	Department of Physics, Zhongshan University, Guangzhou 510275, China\\
        }
       
\footnotetext{Corresponding author: Simon Catterall, 
email: {\tt smc@physics.syr.edu}}
\end{center}

\begin{abstract}

We report on a study of the supersymmetric anharmonic oscillator
computed using a euclidean lattice path integral. Our numerical work
utilizes a Fourier accelerated hybrid Monte Carlo scheme to sample
the path integral. Using this we are able to measure massgaps
and check Ward identities to a precision of better than one percent.
We work with a non-standard lattice action which we show
has an {\it exact} supersymmetry for
arbitrary lattice spacing in the limit of zero interaction
coupling. For the interacting model we show that supersymmetry
is restored in the continuum limit without fine tuning. 
This is contrasted with the situation in which a `standard'
lattice action is employed. In this case supersymmetry is not
restored even in the limit of zero lattice spacing.
Finally, we show how a minor
modification of our action leads to an {\it exact}, local
lattice supersymmetry even in the presence of interaction.

\end{abstract}

\section*{Introduction}

Supersymmetry is thought to be a crucial ingredient in any theory which
attempts to unify the separate interactions contained in the standard model
of particle physics. Since low energy physics is manifestly not
supersymmetric it is necessary that this symmetry be broken at some
energy scale. Issues of spontaneous symmetry breaking have proven 
difficult to address in perturbation theory and hence one is 
motivated to have some non-perturbative method for investigating
such theories. The lattice furnishes such a framework. Unfortunately,
supersymmetry being a spacetime symmetry is explicitly broken by the
discretization procedure and it is highly non-trivial problem to
show that it is recovered in the continuum limit.

One manifestation of this problem is the usual doubling problem of
lattice fermions - the naive fermion action in $D$ dimensions
possesses not one but
$2^D$ continuum-like
modes. These extra modes persist in the continuum limit and
yield an immediate conflict with supersymmetry requiring as it does 
an equality
between boson and fermion degrees of freedom.
As has been noted by several authors \cite{multi} it is possible to circumvent this
problem in a free theory by the addition of a simple Wilson mass term to the
fermion action. This removes the doubles and leads to
a supersymmetric free theory in the continuum limit.
However such a procedure fails when interactions are introduced.
Instead we shall show that the use of a non-standard lattice action
allows
the quantum continuum limit of such an interacting theory to admit
continuum supersymmetry \cite{petcher}. Indeed, we will show that this action has an
exact lattice supersymmetry in the absence of interactions (similar
to that proposed in \cite{biet}) and very small
symmetry breaking effects at non-zero interaction coupling. Finally, we
write down an action for the interacting theory which is
supersymmetric for all lattice spacings.

\section*{Model}

The model we will study contains a real scalar field $x$ and two independent real
fermionic fields $\psi$ and \psib\ defined on a one-dimensional lattice of
$L$ sites with
periodic boundary conditions imposed on both scalar and fermion fields.
\beq
S=\sum_{ij}\frac{1}{2}\left[-x_i D^2_{ij}x_j+
\psib_i\left(D_{ij}+P^\prime_{ij}\right)\psi_j\right]+
\sum_i\frac{1}{2}P_i^2
\label{action}
\eeq
The quantity $P_i$ is defined as
\[P_i=\sum_j K_{ij}x_j+gx_i^3\]
and its derivative is then just
\[P^\prime_{ij}=K_{ij}+3gx_i^2\delta_{ij}\]
The choice of a cubic interaction term in $P(x)$ guarantees unbroken 
supersymmetry \cite{witten} in the continuum.
The matrix $D_{ij}$ is
the symmetric difference operator 
\[D_{ij}=\frac{1}{2}\left[\delta_{j,i+1}-\delta_{j,i-1}\right]\]
and $K_{ij}$ is the Wilson mass matrix 
\[K_{ij}=m\delta_{ij}-\frac{r}{2}\left(\delta_{i,j+1}+
\delta_{i,j-1}-2\delta_{ij}\right)\]
We work in dimensionless lattice units in which $m=m_{\rm phys}a$, $g=g_{\rm
phys}a^2$ and $x=a^{-\frac{1}{2}}x_{\rm phys}$.  

Notice that the boson operator $D^2$ is not the usual lattice Laplacian
$\Box=D_+D_-$ but contains a double corresponding to the extra zero in
$D=\frac{1}{2}(D_+ + D_-)$.
However, the boson action contains now a Wilson mass term 
and so this extra state,
like its fermionic counterpart,
decouples in the continuum limit.
With the further choice $r=1$ the fermion matrix $M=D+P^\prime$ is
almost lower triangular and its determinant can be shown to be
\[{\rm det}(M)=\prod_{i=1}^L\left(1+m+3gx_i^2\right)-1\]
which is
positive definite for $g>0$ and $m>0$. This fact will be
utilized in our numerical
algorithm. Furthermore, the choice $r=1$ removes the doubles {\it completely}
in the free theory and renders the fermion correlators simple
exponentials.

\section*{Simulation}

In order to simulate the fermionic sector we first replace the fermion field
by a bosonic pseudofermion field $\phi$ whose action is just
\[\sum_{ij}\frac{1}{2}\phi_i\left(M^TM\right)^{-1}_{ij}\phi_j\]
This is an exact
representation of the original fermion effective action provided the
determinant of the fermion matrix is positive definite. The resultant (non-local)
action $S\left(x,\phi\right)$ can now be simulated using the
Hybrid Monte Carlo (HMC) algorithm \cite{hmc}. 
In the HMC scheme {\it momentum} fields $(p,\pi)$ conjugate to
$(x,\phi)$ are added and a {\it Hamiltonian} $H$ defined which is just the
sum of the original action plus additional terms depending on the
momenta $H=S+\Delta S$. 
\[\Delta S=\sum_i\frac{1}{2}\left(p_i^2+\pi_i^2\right)\]
On integrating out the momenta it is clear that this partition function is (up to a
constant) identical to the original one. The augmented system $(x,p,\phi,\pi)$
is now naturally associated with some classical dynamics depending
on an auxiliary
time variable $t$
\begin{eqnarray*}
\frac{\partial x}{\partial t}&=&p\\
\frac{\partial p}{\partial t}&=&-\frac{\partial H}{\partial x}\\
\frac{\partial \phi}{\partial t}&=&\pi\\
\frac{\partial \pi}{\partial t}&=&-\frac{\partial H}{\partial\phi}
\end{eqnarray*}

If we introduce a finite time step $\Delta t$ we may simulate this classical
evolution and produce a sequence of configurations $(x(t),\phi(t))$. If
$\Delta t=0$ then $H$ would be conserved along such a trajectory. In practice
$\Delta t$ is finite and $H$ is not exactly conserved. However a finite
length of such an approximate trajectory can still be used as a
{\it global} move on the fields $(x,\phi)$ which may then be subject to
a Metropolis step based on $\Delta H$. Provided the classical dynamics is
reversible and care is taken to ensure
ergodicity the resulting move satisfies detailed balance and hence this
dynamics will provide a simulation of the
original partition function. The reversibility criterion can
be satisfied by using a leapfrog integration scheme and ergodicity
is taken care of by drawing new momenta from a Gaussian
distribution after each such trajectory. 

If we introduce bosonic and pseudofermionic forces
\beq
F_i(t)=-\frac{\partial H(t)}{\partial x_i}
\eeq
and
\beq
{\mathcal F}_i(t)=-\frac{\partial H(t)}{\partial \phi_i},
\eeq
the resultant evolution equations look like
\begin{eqnarray}
x_i(t+\Delta t)&=&x_i(t)+\Delta t p_i(t)-\frac{(\Delta t)^2}{2}F_i(t)\nonumber\\
\phi_i(t+\Delta t)&=&\phi_i(t)+\Delta t \pi_i(t)-\frac{(\Delta
t)^2}{2}{\mathcal F}_i(t)\nonumber\\
p_i(t+\Delta t)&=&p_i(t)+\frac{\Delta t}{2}\left(F_i(t)+
F_i(t+\Delta
t)\right)\nonumber\\
\pi_i(t+\Delta t)&=&\pi_i(t)+\frac{\Delta t}{2}\left({\mathcal F}_i(t)+
{\mathcal F}_i(t+\Delta t)\right)
\label{update}
\end{eqnarray}
The force terms are then given in terms of a vector $s_i$ which is a solution of
the (sparse) linear problem
\[\left(M^TM\right)_{ij}s_j=\phi_i\]
\begin{eqnarray*}
\frac{\partial H}{\partial x_i}&=&-D^2_{ij}x_j+P^\prime_{ij}P_j+6gx_is_iM_{ij}s_j\\
\frac{\partial H}{\partial \phi_i}&=&s_i
\end{eqnarray*}
In order to reduce the effects of critical slowing down we have chosen to
perform this update in momentum space using FFTs and
a momentum dependent time step.
Thus, for example, the lattice field $x_m, (m=0\ldots L-1$) can be
expanded as
\[x_m=\sqrt{\frac{1}{L}}\sum_n \overline{x}_n e^{i\frac{2\pi mn}{L}}\]
where $\overline{x}_n$ is the Fourier amplitude with wavenumber $\frac{2\pi
}{L}n$ ($n=0\ldots L-1$) and the Fourier
amplitudes $\overline{x}_n$ are updated using equations \ref{update} with
$\Delta t=\Delta t(n)$.
For the boson field $x$ we use $\Delta t=\tau_B$ where
\[
\tau_B(n)=\epsilon\frac{\left(m_{\rm eff}+2r\right)}
        {\sqrt{\sin^2{\frac{2\pi n}{L}}+\left(m_{\rm eff}+2r\sin^2{\frac{2\pi n}{2L}}\right)^2}}\]
For the pseudofermion field update we use the inverse
function $\Delta t=\tau_F=\frac{1}{\tau_B}$. With these choices (and
$m_{\rm eff}=ma$) it
is simple to show that the $g=0$ theory suffers no critical slowing down
- all modes are updated at the same rate independent of their wavelength.
By setting $m_{\rm eff}$ at the approximate position of the massgap in the
interacting case we have found very substantial reductions in the
autocorrelation time for the two point functions of the theory.
In practice we set $\epsilon\sim 0.1$ and the number of leapfrog integrations
per trajectory at $N_{\rm leap}=10$.

\section*{Correlation functions}

We have measured the following correlators
\[G^B_{ij}=\left<x_ix_j\right>\]
and
\[G^F_{ij}=\left<s_jM_{ik}s_k\right>\]
It can be shown that the latter is simply an estimator for the original
fermion correlator $<\overline{\psi}_i\psi_j>$. 

\begin{figure}[hb]
\begin{center}
\includegraphics[width=11cm]{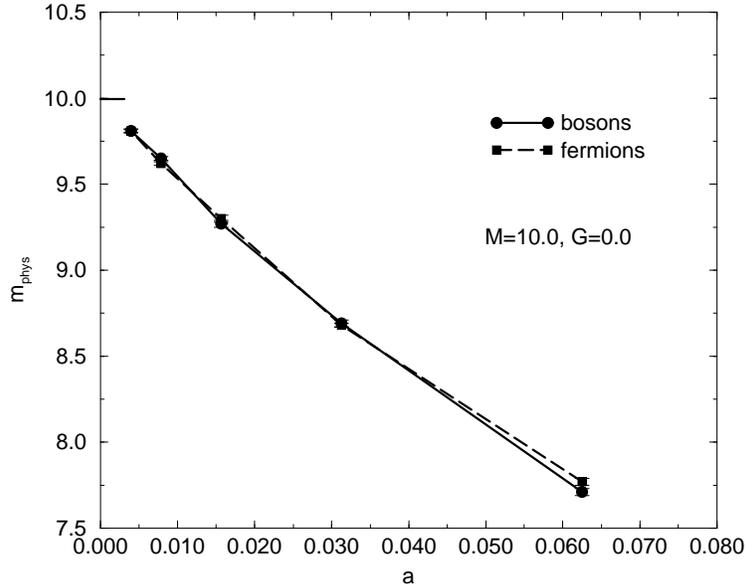}
\end{center}
\caption{\label{fig1} Boson and Fermion masses vs. lattice spacing $a$ at
$m_{\rm phys}=10.0$, $g_{\rm phys}=0.0$}
\end{figure} 	   
    
In figure~\ref{fig1} we show first the results of a simulation of this model
for $g_{\rm phys}=0.0$ and $m_{\rm phys}=10.0$. 
The data set consists of $10^6$ Fourier accelerated
HMC trajectories. The plot shows both boson and fermion
massgaps, extracted from a simple exponential fit to the correlators over
the first $L/4$ timeslices, as a function of the lattice spacing $a=1/L$. 
Notice that boson and fermion masses while
receiving large O(a) systematic errors (due to the
Wilson term) are degenerate within statistical
errors.  We see furthermore that as ${\rm a}\to 0$ the 
common massgap approaches the correct
continuum value. As we shall see later the free action has an {\it exact}
supersymmetry at finite lattice spacing which is responsible for
the boson/fermion degeneracy. 

\begin{figure}[htb]
\begin{center}
\includegraphics[width=11cm]{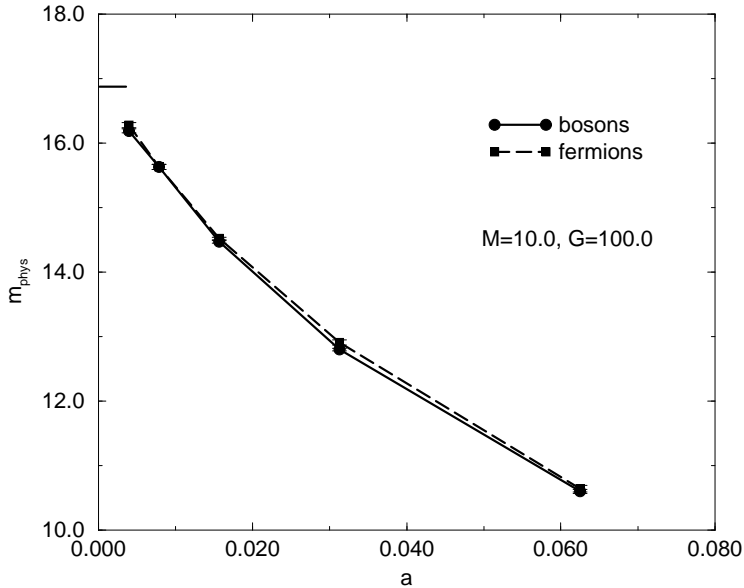}
\end{center}
\caption{\label{fig2} Boson and Fermion masses vs. lattice spacing $a$ at
$m_{\rm phys}=10.0$, $g_{\rm phys}=100.0$}
\end{figure}
\begin{figure}[htb]
\begin{center}
\includegraphics[width=11cm]{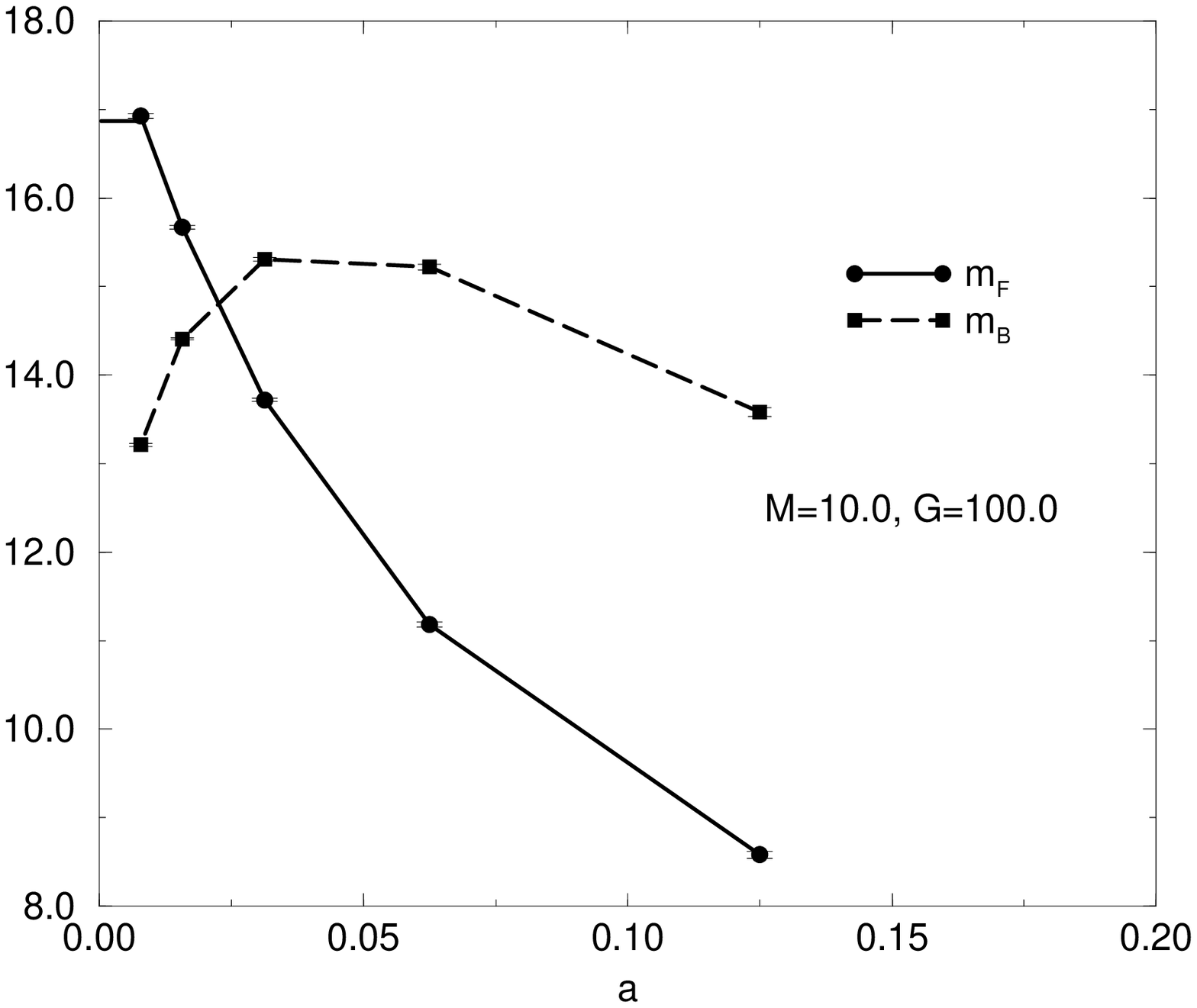}
\end{center}
\caption{\label{fig3} Boson and Fermion masses vs. lattice spacing $a$ at
$m_{\rm phys}=10.0$, $g_{\rm phys}=100.0$ computed from `naive' action}
\end{figure}

   
We have also examined the massgaps at non-zero coupling. Figure~\ref{fig2}
shows the same plot for $m_{\rm phys}=10.0$ and $g_{\rm phys}=100.0$. The massgaps are
also listed in Table~\ref{table1}. 
\begin{table}
\label{table1}
\begin{center}
\begin{tabular}{|@{\hspace{0.5cm}}l@{\hspace{0.5cm}}|@{\hspace{1.7cm}}l@{\hspace{1.7cm}}|@{\hspace{1.7cm}}l@{\hspace{1.7cm}}|}
\hline
L  & $m_B$      &$m_F$      \\\hline
16 & $10.60(3)$ & $10.64(5)$ \\\hline
32 & $12.80(2)$ & $12.91(4)$ \\\hline
64 & $14.47(2)$ & $14.52(2)$ \\\hline
128 & $15.63(4)$ & $15.63(4)$ \\\hline
256 & $16.19(3)$ & $16.28(4)$ \\\hline
\end{tabular}
\caption{Boson and Fermion massgaps for $g/m^2=1$ versus inverse lattice
spacing}
\end{center}
\end{table}
The data set
consists of $10^6$ trajectories again using lattice sizes $L=16-256$.
The effective dimensionless
expansion parameter is $g/m^2$ so this corresponds to a regime of
{\it strong} coupling. Remarkably, the boson and fermion masses are
again degenerate within statistical errors O($0.5$\%) and flow
as ${\rm a}\to 0$ to the correct continuum limit (the latter can be
computed easily using Hamiltonian methods and yields $m_{\rm cont}=16.87$)). 
That this result is nontrivial
can be seen when we compare it to the result of a `naive' discretization of
the continuum action using $\Box$ in place of $D^2$ and the usual
Wilson action for the fermions -- figure~\ref{fig3}.
 
In this case the mass plot looks very different. At large
lattice spacing the extracted massgaps differ widely -- the fermion
having O(a) errors while the boson is much smaller
(it varies as O(${\rm a}^2$) at $g=0$).
Initially they appear to approach each other as ${\rm a}\to 0$ but the
two curves depart for fine lattice spacing and do not approach the correct
continuum limit - the quantum continuum limit is {\it not} supersymmetric.
Thus naive discretizations of the continuum action will break
supersymmetry {\it irreversibly} even in theories such as quantum mechanics which
have no divergences. At minimum it would be necessary to tune
parameters to obtain a supersymmetric continuum limit.
In comparison the numerical results
of figure~\ref{fig2} indicate that supersymmetry breaking effects, if present,
are very small. We examine this more carefully next.

\section*{Supersymmetry}

Motivated by the form of the continuum supersymmetry transformations for this
model consider the following two lattice transformations
\begin{eqnarray}
\delta_1 x_i &=& \overline{\psi_i}\xi\nonumber\\
\delta_1 \psi_i &=& \left(D_{ij}x_j-P_i\right)\xi\nonumber\\
\delta_1 \overline{\psi_i}&=&0
\label{susy1}
\end{eqnarray}
and
\begin{eqnarray}
\delta_2 x_i& = &\psi_i \overline{\xi}\nonumber\\
\delta_2 \psi_i& =& 0\nonumber\\
\delta_2 \overline{\psi_i}&= &\left(D_{ij}x_j+P_i\right)\overline{\xi}
\label{susy2}
\end{eqnarray}
where $\xi$ and $\overline{\xi}$ are independent anti-commuting parameters. The existence of two
such symmetries reflects the $N=2$ character of the
continuum supersymmetry. 
If we perform the variation corresponding to the first of these 
(eqn.~\ref{susy1})
we find
\beq
\delta_1 S=\sum_i\psib_i\xi\left(P^\prime_{ij}D_{jk}x_k-D_{ij}P_j\right)
\label{delta1}
\eeq
The expression corresponding to the second transformation eqn.~\ref{susy2}
is similar
\beq
\delta_2 S=\sum_i\overline{\xi}\psi_i\left(P^\prime_{ij}D_{jk}x_k-D_{ij}P_j
\right)
\label{delta2}
\eeq
In the continuum limit $a=0$ the difference operators become derivatives
and the term inside the brackets is zero - this is the statement of
continuum supersymmetry. Notice, for $g=0$ and $a\ne 0$ this term is still
zero - the classical free lattice action is also supersymmetric.
However this term is non-zero for finite spacing and non-zero
interaction coupling -
the classical lattice action breaks supersymmetry. Since we use the symmetric difference operator this
breaking will be ${\rm O}(gL^{-2})$. From the point of view of
a continuum limit such a breaking would not be important - since the theory
contains no divergences all non-supersymmetric terms induced in the quantum
effective action will have couplings that vanish as $a\to 0$. Indeed, as was shown
explicitly in \cite{petcher}, the two-dimensional Wess-Zumino model
has a supersymmetric continuum limit when regulated in
this way. For a lattice of size $L=16$ and $g_{\rm phys}=100.0$ we would
expect symmetry breaking terms to be suppressed by a factor of $g_{\rm phys}/L^4=0.002$.
This is consistent with what we see in the massgaps.

To verify these conclusions we have studied the approximate Ward identities
which follow from the lattice transformations eqn.~\ref{susy1} and eqn.~\ref{susy2}.

\section*{Ward identities}

The Ward identities corresponding to these approximate symmetries can be derived in the 
usual way. First consider the partition function with external sources
\[Z\left(J,\theta,\overline{\theta}\right)=\int Dx D\psi D\overline{\psi}
e^{-S+\sum J.x+\overline{\theta}.\psi+\theta.\overline{\psi}}\]
Perform a lattice supersymmetry transformation, for example, eqn.~\ref{susy1}.
The action $S$ varies as in eqn.~\ref{delta1},
and the integration measure is invariant while the
source terms vary. Since the partition function does not change (the
transformation can be
viewed as a change of variables) we find
\beq
0=\delta Z=\int Dx D\psi D\overline{\psi}e^{-S}\left(\sum J.\delta
x+\overline{\theta}.\delta\psi+\theta.\delta\overline{\psi}+\alpha.\psib\right)
\eeq
where $\alpha_i=\left(P^\prime_{ij}D_{jk}x_k-D_{ij}P_j\right)$
Furthermore, any number of derivatives with respect to the sources evaluated
for zero sources will also vanish. For the first supersymmetry equation \ref{susy1}
this yields a set of identities connecting different correlation functions.
The first non-trivial example is
\[
\left<\overline{\psi_i}\psi_j\right>+\left<(D_{jk}x_k-P_j)x_i\right>+
\left<\alpha_k\psib_k x_i\psi_j\right>=0
\]
The last term represents the symmetry breaking term which may be rewritten
as $<(M^T)^{-1}_{jk}\alpha_k x_i>$. Since $\alpha_k$ is a vector with
random elements each of mean zero and suffering fluctuations O($g/L^2)$) we might
expect that this term contributes rather a small correction to the
Ward identity. In this spirit we will neglect it at
this point and see if the predictions are
substantiated by the results of the simulation. 
It is important to notice that
this correction to the naive Ward identity is finite (quantum
mechanics) and multiplied
by $1/L^2$ and
consequently the continuum limit is {\it guaranteed} to possess supersymmetry.
The same will be true for any approximate Ward identity we care to
construct.
 
A second Ward identity may be
derived corresponding to the second (approximate)
supersymmetry equation \ref{susy2} we find
\[ 
\left<\psi_i\overline{\psi_j}\right>+\left<(D_{jk}x_k+P_j)(x)x_i\right>=0
\]
More conveniently we can add and subtract these equations to yield the
relations
\beq
G^F_{ij}-G^F_{ji}=-\left<2D_{jk}x_kx_i\right>
\label{w2}
\eeq
and
\beq
G^F_{ij}+G^F_{ji}=\left<2P_jx_i\right>
\label{w1}
\eeq
Translation invariance on the lattice implies $G^F_{ij}=G^F(t)$ and
$G^F_{ji}=G^F(L-t)$ where $t=(j-i)$.
We can check that these two Ward identities are satisfied numerically by
forming the two (distance dependent) quantities $w_1$ and $w_2$ which are
defined by
\beq
w_1\left(t\right)=\frac{G^F_{ij}+G^F_{ji}-2\left<x_iP_j\right>}
{2\left<x_iP_j\right>}
\eeq
\beq
w_2\left(t\right)=\frac{G^F_{ij}-G^F_{ji}+2\left<x_iD_{jk}x_k\right>}
{2\left<x_iD_{jk}x_k\right>}
\eeq

\begin{figure}[htb]
\begin{center}
\includegraphics[width=11cm]{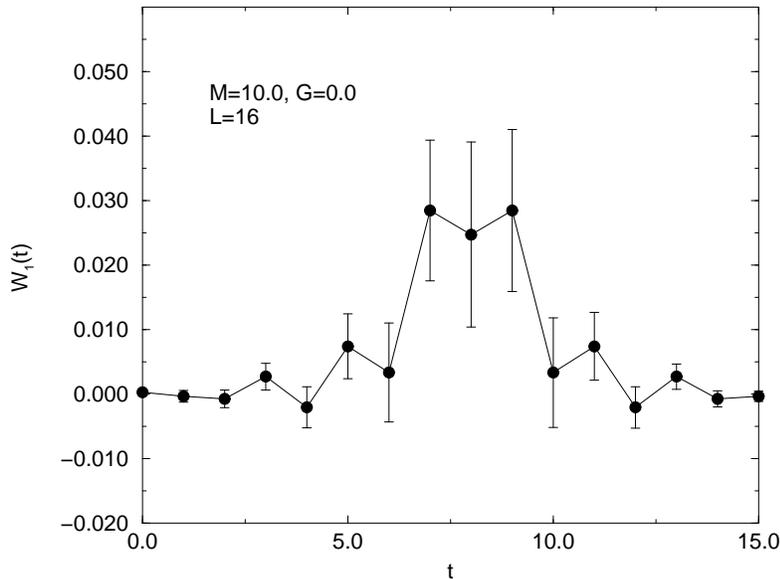}
\end{center}
\caption{\label{fig4} $w_1(t)$ vs. $t$ for $L=16$, $m_{\rm phys}=10.0$
 and $g_{\rm phys}=0.0$}
\end{figure} 

\begin{figure}[htb]
\begin{center}
\includegraphics[width=11cm]{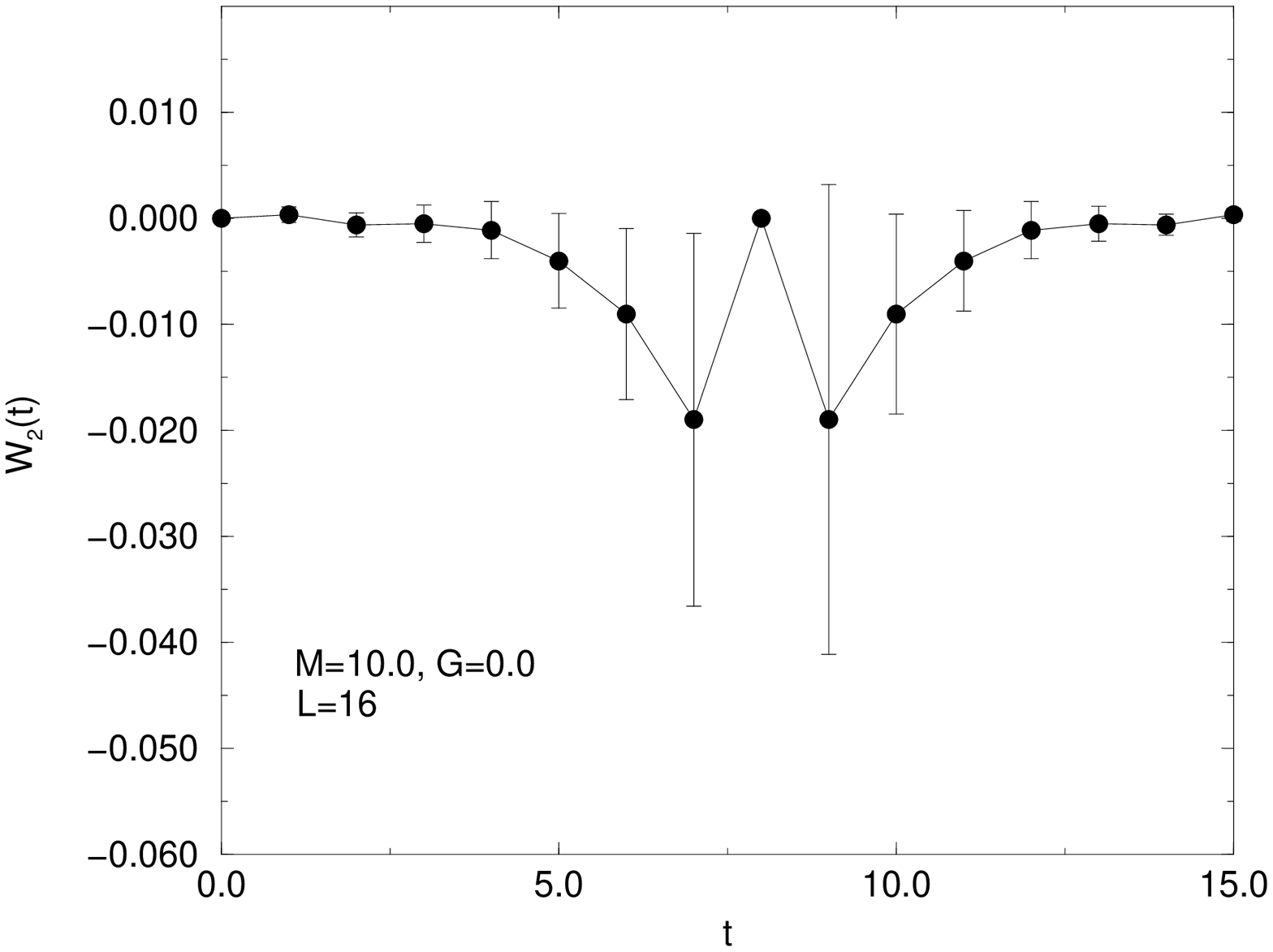}
\end{center}
\caption{\label{fig5} $w_2(t)$ vs. $t$ for $L=16$, $m_{\rm phys}=10.0$
 and $g_{\rm phys}=0.0$}
\end{figure} 

These are shown in figure~\ref{fig4} and figure~\ref{fig5} for a lattice of
size $L=16$ at $g_{\rm phys}=0.0$. In this case we expect
the symmetry to be exact and indeed we see that
the Ward identities are satisfied
within statistical accuracy. 

\begin{figure}[htb]
\begin{center}
\includegraphics[width=11cm]{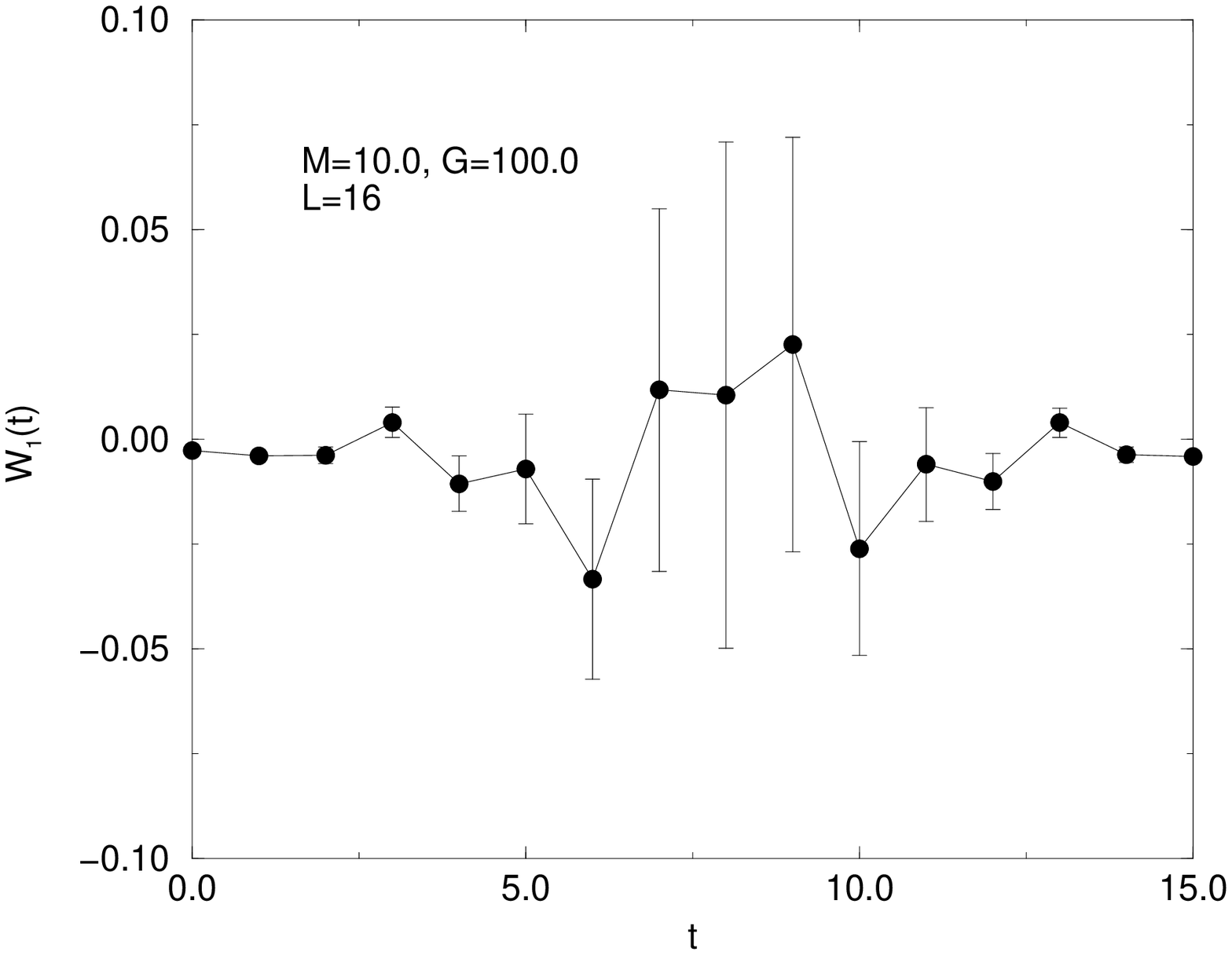}
\end{center}
\caption{\label{fig6} $w_1(t)$ vs. $t$ for $L=16$,$m_{\rm phys}=10.0$ and
$g_{\rm phys}=100.0$}
\end{figure}

\begin{figure}[htb]
\begin{center}
\includegraphics[width=11cm]{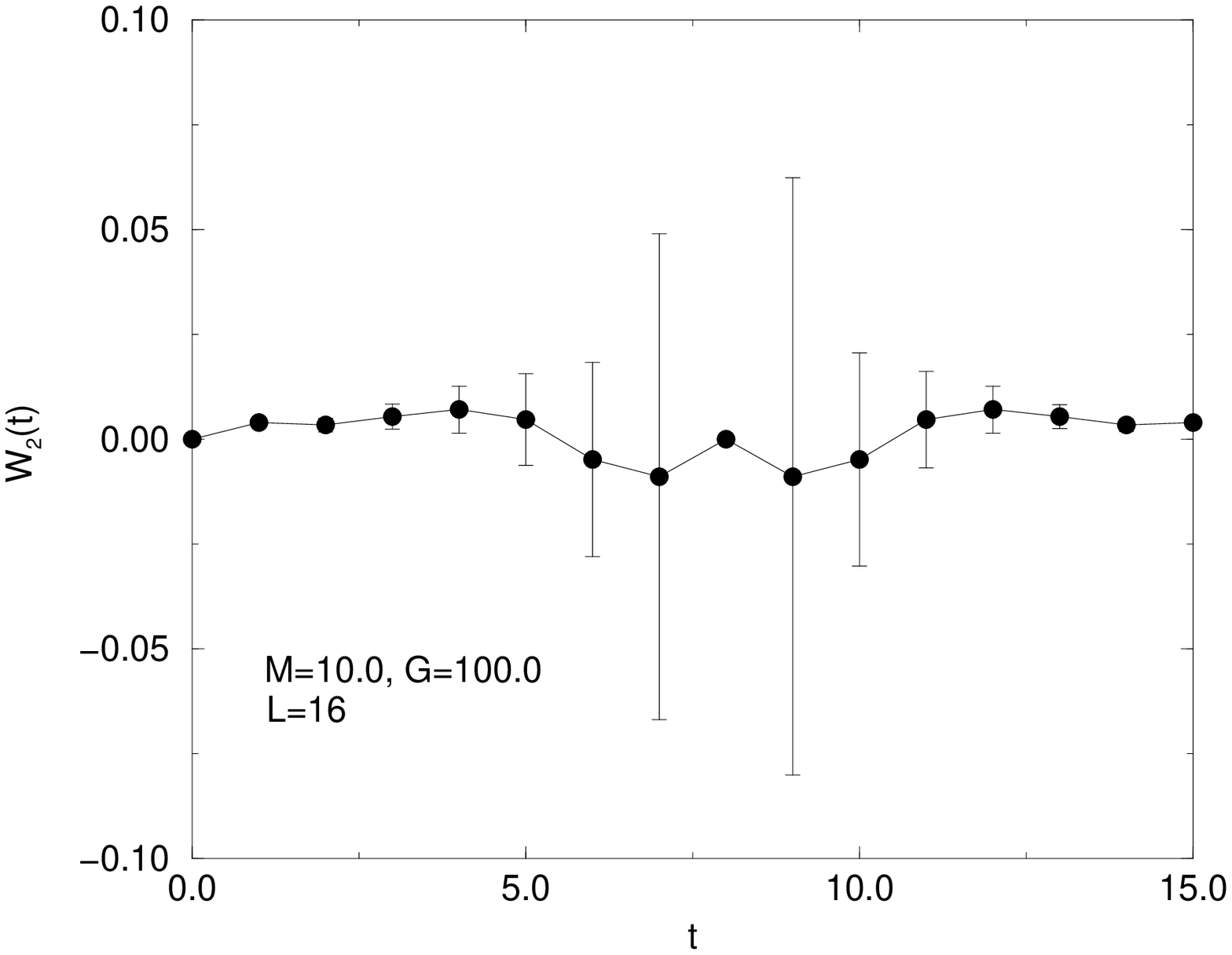}
\end{center}
\caption{\label{fig7} $w_2(t)$ vs. $t$ for $L=16$, $m_{\rm phys}=10.0$ and
$g_{\rm phys}=100.0$}
\end{figure}

Figures~\ref{fig6} and \ref{fig7} show plots of $w_1$ and $w_2$ 
for a lattice of size $L=16$ at $g_{\rm phys}=100.0$. It is
clear that within our statistical error (on the order of a few percent for
these quantities) we are again not sensitive to the SUSY breaking terms 
and the continuum Ward identities are satisfied. 

\section*{Discussion and Conclusions}

We have performed a numerical study of the lattice supersymmetric anharmonic
oscillator computed using path integrals. This is
essentially a one-dimensional version of the Wess-Zumino model.
We have utilized a lattice discretization
which preserves two exact supersymmetries in the free theory. We are
able to show that the interacting theory flows to a supersymmetric fixed point in the
zero lattice spacing limit without fine tuning. This is to
be contrasted with
naive discretizations of the continuum action which require fine tuning
to recover supersymmetry in the continuum limit. 

Furthermore, we have estimated the magnitude of supersymmetry breaking
at O($g/L^2$) which is typically smaller than one percent even at strong
coupling and for coarse lattices. Thus the lattice simulations are, in
practice, very close to the supersymmetric fixed point.
We have checked the first two
non-trivial 
Ward identities following from this (approximate) invariance. Our numerical
results 
place an upper bound on the magnitude of symmetry breaking
corrections which is consistent with this estimate.
 
It is tempting to try to interpret the numerical results as evidence of
an {\it exact} lattice supersymmetry even in the presence of
interactions. Using the antisymmetry of the derivative operator it is easy to
show that the symmetry breaking term~\ref{delta2} can be rewritten
\[\delta_2 S=\delta_2\sum x_iD_{ij}P_j\]
Thus $S_{\rm new}=S-\sum x_iD_{ij}P_j$ will be {\it exactly} invariant under
the second lattice supersymmetry transformation even for
non zero interaction. Notice that the presence of an extra
minus sign prevents this new action from having a second invariance 
corresponding to the first supersymmetry~\ref{delta1}. This invariant action can
be rewritten in the form
\[S_{\rm new}=\sum_{ij}\frac{1}{2}\left(D_{ij}x_j+P_i\right)
\left(D_{ij}x_j+P_i\right)+
\sum_{ij}\psib_i\left(D_{ij}+P^\prime_{ij}\right)\psi_j\]
This allows us to identify the Nicolai map for the model \cite{nic1}. 
The latter is the non-trivial transformation $x_i\to \xi_i$
which maps the boson action to a free field
form and whose Jacobian simultaneously cancels the fermion determinant.
Here we see it explicitly
\[\xi_i=D_{ij}x_j+P_i\]
It has previously been pointed out that the identification of such a map
may be used to help find lattice supersymmetric actions \cite{nic2}
and \cite{becc}. This
quantum mechanics model furnishes a concrete example - the lattice 
action which
admits the Nicolai map is invariant under a transformation which
interchanges bosonic and fermionic degrees of freedom. In the continuum
limit this lattice action approaches its continuum counterpart and the
transformation reduces to a continuum supersymmetry
transformation. The presence of one exact supersymmetry is already
enough to guarantee
vanishing vacuum energy and boson/fermion mass degeneracy for the lattice
theory. 
   
It would be interesting to extend these calculations to the two-dimensional
Wess-Zumino model and verify {\it non-perturbatively} the results
derived perturbatively in \cite{petcher}.

\section*{Acknowledgements}
Simon Catterall was supported in part by DOE grant DE-FG02-85ER40237. We would
like to thank Yoshio Kikukawa for drawing our attention to refs~\cite{nic1}
and \cite{nic2}.


\begin{thebibliography}{9}
\bibitem{multi} Hiroto So and Naoya Ukita, Phys.Lett. B457 (1999) 314\\
Tatsumi Aoyama and Yoshio Kikukawa, Phys.Rev. D59 (1999)
\bibitem{petcher} M. Goltermaan and D. Petcher, Nucl. Phys. B319 (1989) 307.
\bibitem{biet} W. Bietenholz, Mod. Phys. Lett A14 (1999) 51.
\bibitem{witten} E. Witten, Nucl. Phys. B185 (1981) 513.
\bibitem{hmc} S. Duane, A. Kennedy, B. Pendleton and D. Roweth, Phys. Lett.
B195B (1987) 216.
\bibitem{nic1} H. Nicolai, Phys. Lett. B89 (1980) 341.
\bibitem{nic2} N. Sakai and M. Sakamoto, Nucl. Phys. B229 (1983) 173.
\bibitem{becc} M. Beccaria, G. Curci and E. D'Ambrosio, Phys. Rev. D58 (1998)
065009.
\end{thebibliography}
\end{document}